\documentstyle[multicol,aps,prl,graphics,epsfig]{revtex}
\begin{document}
\title{Exploring Hilbert Space: accurate characterisation of quantum
information}
\author{A. G. White$^{1}$, D. F. V. James$^{2}$, W. J. Munro$^{1,3}$,
and P. G. Kwiat$^{2,4}$}
\address{$^{1}$ Department of Physics, University of Queensland,
Brisbane, Queensland 4072, AUSTRALIA}
\address{$^{2}$ Theory Division, T-4; Physics Division, P-23;
Los Alamos National Laboratory, Los Alamos, New Mexico 87545, USA}
\address{$^{3}$ Maths, Cryptography and Security Group, Hewlett-Packard
Laboratories, Bristol BS34 8QZ, UK}
\address{$^{4}$ Department of Physics, University of Illinois, 
Urbana-Champaign, Illinois 61801, USA}
\draft
\maketitle
\begin{abstract}
We report the creation of a wide range of quantum states with 
controllable degrees of entanglement and entropy using an optical 
two-qubit source based on spontaneous parametric downconversion.  The 
states are characterised using measures of entanglement and entropy 
determined from tomographically determined density matrices.  The 
Tangle-Entropy plane is introduced as a graphical representation of 
these states, and the theoretic upper bound for the maximum amount of 
entanglement possible for a given entropy is presented. Such a combination 
of general quantum state creation and accurate characterisation is an 
essential prerequisite for quantum device development.
\end{abstract}

\begin{multicols}{2}
Quantum information (QI) - the application of quantum mechanics to 
problems in information science such as computation and communication 
- has led to a renewed interest in fundamental aspects of quantum 
mechanics.  In particular much attention has been focussed on the role 
of {\it entanglement\/}, the non-classical correlation between 
separate quantum systems, particularly between two-level systems or 
{\it qubits\/} (quantum bits).  Entanglement, along with the degree of 
order, or {\it purity\/}, determines the utility of a given system for 
realising various QI protocols.  A key goal of QI is the experimental 
realisation of complex quantum algorithms, e.g., Shor's algorithm, 
which allows efficient factoring of composite integers \cite{Shor}: 
recent research indicates that while entanglement is necessary to 
execute Shor's algorithm, pure states are not \cite{Parker}.

There is currently a global effort to manufacture two-qubit gates, 
since any quantum algorithm can be implemented by a combination of 
single qubit rotations and such gates (which produce the necessary 
entanglement) \cite{DiVincenzo}.  These are fully characterised only 
when both the gate states and its dynamics have been accurately 
measured, which requires a tunable source of two-qubit quantum states 
and a method of completely measuring the output states \cite{Chuang}.  
No system to date has fulfilled these criteria.  Here we report an 
optical two-qubit source that produces a wide range of quantum states 
with controllable degrees of purity and entanglement, and fully 
characterise these states by quantum tomography.  This source is also 
suitable for exploring alternative paradigms: (1) where quantum 
algorithms are implemented via single qubit rotations, Bell-state 
measurements, and a pre-determined set of entangled states (which may 
or may not need to be pure) \cite{Gottesman}; (2) scaleable 
linear-optics quantum computation, where entanglement occurs as a 
result of measurement \cite{KLM}.

Quantum states of N qubits can be represented by a vector existing in 
a $2^{\rm{N}}$-dimensional Hilbert space.  This is the ``space of 
possibilities'', and represents all possible physical combinations of 
qubits for a system.  To date the states generated in QI experiments 
have clustered around two distinct limits in Hilbert space: 1) highly 
entangled systems with high order 
\cite{cavQED,dense,teleport1,ion1,crypto2}; 2) completely unentangled 
systems with very little order \cite{NMR1}.  The lack of entanglement 
in the latter case \cite{notentang} has raised the question of what 
properties are actually required for quantum information protocols, 
and highlights that to date, the ``domain of mixed states between 
these two extremes [pure vs.  completely mixed] is incredibly big and 
largely unexplored.''  \cite{PTKimble}.  We experimentally explore 
this unmapped region, and introduce a theoretical upper bound for the 
maximum amount of entanglement possible for a given purity.  A variety 
of measures exist for quantifying the degrees of disorder and 
entanglement, all of which are functions of the system density matrix.  
For our experimental system, the density matrix can now be obtained 
via quantum tomography \cite{2Tomo}, allowing these measures to be 
applied.  In this paper we will use the {\it tangle\/}, T, to quantify 
the degree of entanglement, and the {\it linear entropy\/}, 
$\rm{S}_{\rm{L}}$, to quantify the degree of disorder \cite{footnote}.

We obtain our quantum states via spontaneous downconversion, where a 
pump photon passed through a nonlinear crystal is converted into a 
pair of lower energy photons.  We use the polarisation state of the 
single photons as our qubits, and measure in coincidence, thus 
obtaining the reduced density matrix (it only describes the 
polarisation component of the state) of the two-photon contribution 
\cite{comment}.  Figure 1 is a schematic of the experimental system; a 
detailed description is given elsewhere \cite{Ultrabright}.  Briefly, 
the BBO crystals produce pure-state pairs of photons that can be tuned 
between the separable and maximally entangled limits by adjusting the 
pump polarisation \cite{2Tomo}.  The parity and phase of the entangled 
states are selected via the ``state selector'' half-wave plates, and 
the photons are analysed using adjustable quarter- and half-wave 
plates (QWP \& HWP) and polarizing beamsplitters (PBS), which enable 
polarisation analysis in any basis.  (For tomography, 16 different 
coincidence bases are required \cite{2Tomo,TomoErr}).  The photons are 
passed via suitable optics to single photon counters, whose outputs 
are recorded in coincidence.

To change the entropy it is necessary to introduce decoherence into 
the polarisation degree of freedom, which can be done either spatially 
or temporally.  Decoherence can occur when the phase, $\phi$, of the 
entangled state (e.g. $|\rm{HH}\rangle+e^{i \phi}|\rm{VV}\rangle$) 
varies rapidly over a small spatial extent, i.e. smaller than the 
collection apertures.  We achieved this by the introduction of BBO 
crystals (3 mm thick) into the downconversion beams, cut so that their 
optic axes were at an angle of $49^{\circ}$ to the beam.  These 
introduce a highly direction-dependent phase shift in the 
downconverted photons: due to the intrinsic spread of photon momentum 
in downconversion, and the high birefringence of BBO, after the 
decoherer the phase of the entanglement is very finely fringed 
compared to the collection aperture.  Figure 2a shows that with BBO in 
only one arm (optic axis at 45$^{\circ}$), the resultant state is 
partially mixed and completely unentangled, ($\rm{S_{L}}$,T)=($0.66 
\pm 0.03$, $0.00 \pm 0.00$); adding a BBO crystal to the remaining 
arm, (optic axis at 0$^{\circ}$) generates a fully mixed state, 
($\rm{S_{L}}$,T)=($1.00 \pm 0.01$, $0.00 \pm 0.00$).

As spatially-based decoherence completely destroys entanglement, it is 
not a suitable technique for exploring Hilbert space.  In contrast, 
temporally-based polarisation decoherence allows entanglement to 
survive.  It is achieved by imposing a large relative phase delay, 
longer than the coherence length of the light, between two orthogonal 
polarisations.  In practice this is realised by introducing a 10mm 
thick quartz crystal in each arm, with optic axes vertical and 
perpendicular to the beam.  The detected photons have a coherence 
length of ~140 wavelengths (set by the 5 nm interference filters): 10 
mm of quartz delays the phase velocity of the horizontally polarised 
light by this amount (relative to the vertical).  Viewed differently, 
the quartz entangles the phase to the photon frequency, which is then 
traced over \cite{Berglund,comment2}.  Thus, a single photon linearly 
polarised at 45$^{\circ}$ would exit the quartz crystal strongly 
depolarised, i.e. in a mixed state; similarly, a pair of photons, each 
at 45$^{\circ}$, would exit two such crystals in a mixed state.  With 
entangled states, however, the situation is more subtle.  Certain 
kinds of entangled states are immune to collective decoherence, whilst 
others exhibit strong decoherence - the former comprise {\it 
decoherence free subspaces\/} \cite{DFS1}.  Due to the energy 
entanglement of the photons, and the alignment of our decoherers, in 
our system the state $(|\rm{HH} \rangle+|\rm{VV} \rangle)/\sqrt{2}$ is 
decoherence free \cite{Berglund,DFS2}.  The function of the state 
selector is to continuously tune this state towards another 
maximally-entangled state, one that is not decoherence-free (e.g. 
$(|\rm{HV} \rangle+|\rm{VH} \rangle)/\sqrt{2}$).

Figures 2b \& 2c show a range of density matrices generated via this 
method.  In Figure 2b, selector waveplate HWP1 was fixed at 
$0^{\circ}$ and HWP2 varied by the angle indicated; Figure 2c shows a 
similar series, except this time starting with a non-maximally 
entangled state.  There are 16 parameters in the density matrix, too 
many for easy assimilation.  To see how much of Hilbert Space we are 
accessing with these states, we use the tangle and linear entropy 
measures to construct a characteristic plane as a succinct, compact 
way of representing the salient features of a quantum state.  In this 
plane (Figure 3), a pure, unentangled state lies at the origin 
($\rm{S_{L}}$,T)=(0,0); a pure, maximally-entangled state in one 
corner ($\rm{S_{L}}$,T)=(0,1); and a maximally-mixed, unentangled 
state in the corner diagonally opposite ($\rm{S_{L}}$,T)=(1,0).  A 
maximally-entangled, maximally-mixed state ($\rm{S_{L}}$,T)=(1,1) is 
obviously impossible.  As indicated above, previous QI experiments 
have generated states either near the tangle axis ($\rm{S_{L}} \sim 
0$, $0 \leq \rm{T} \leq 1$) (cavity QED, ion and photon experiments) 
or at the maximally mixed point ($\rm{S_{L}} \sim 1-10^{-4}$ to 
$1-10^{-6}$, $\rm{T}=0$) (high-temperature NMR experiments).  In 
Figure 3, we plot the linear entropy and tangle values determined from 
a range of our measured density matrices, including those shown in 
Figure 2 and from \cite{Distill}.  Two sources of experimental 
uncertainty were considered: statistical uncertainties ($\sqrt{N}$, 
where $N$ is the count), which range from 1-4\% for our count rates; 
and settings uncertainties, due to the fact that the analysers can 
only be set with an accuracy of $\pm 0.25^{\circ}$.  The combination 
of these effects led to the uncertainties as shown - a full derivation 
of their calculation is lengthy and given elsewhere \cite{TomoErr}.  
The heavy black line in Figure 3 represents the (S,T) values of Werner 
states ($\hat{\rho}_{W}=\lambda \hat{\rho}_{mix}+(1-\lambda) 
\hat{\rho}_{ent}$, where $\hat{\rho}_{mix}$ is a maximally-mixed 
state, $\hat{\rho}_{ent}$ is a pure maximally-entangled state, and 
$0<\lambda<1$) \cite{Werner}.  A wide array of states were created, up 
to and lying on the Werner border.  Interestingly, for linear 
entropies less than 8/9, there exist states with greater Tangle than 
Werner states, the largest of which, the {\it maximal\/} states, lie 
at the boundary of the grey region in Figure 3.  The density matrix 
for these states has the form \cite{Maximal}:
\begin{eqnarray}
\hat{\rho}=\left(
\begin{array}{cccc}
\rm{D}&  0  & 0  & \sqrt{\rm{T}}/2 \\
0 & 1-2 \rm{D} & 0 &  0 \\
0 & 0 & 0 & 0 \\
\sqrt{\rm{T}}/2 &  0  & 0  & \rm{D} \\
\end{array}
\right)
\end{eqnarray}
where $\rm{D}=\sqrt{\rm{T}}/2$ when $\sqrt{\rm{T}} 
\geq 2/3$; and $\rm{D}=1/3$ when $\sqrt{\rm{T}} < 2/3$.  Only states 
with tangle and linear entropy that fall on or under this boundary are 
physically realisable.  Using the current scheme of 2 decohering 
crystals, it is not possible to create states that lie between the 
Werner and maximal boundaries.  We are currently investigating a 
method to realise generalised arbitrary quantum state synthesis, which 
will allow generation of states with any allowed combination of 
entropy and tangle.

In any experimental system mixture is inevitable - our source enables 
experimental investigation of decoherence-induced effects and issues 
including: entanglement purification \cite{Bennett96}, distillation 
\cite{Distill,Vedral}, concentration \cite{Thew}; decoherence-free 
subspaces \cite{DFS1,DFS2}; and protocols that {\it require\/} 
decoherence \cite{Cleve}.  Since decoherence is controllable in our 
system, it can be used as a testbed for controlled exploration of the 
effect of intrinsic, uncontrollable, decoherence in other architectures 
(e.g. decoherence in a solid-state two-qubit gate \cite{Kane}).

Mixed entangled states also have fundamental ramifications.  
Entanglement, as defined by Schr\"{o}dinger, is essentially a pure 
state concept (resting as it does on the issue of separability) and is 
``...{\it the\/} quintessential feature of quantum mechanics, the one 
that enforces its entire departure from classical lines of thought'' 
\cite{Schroedinger}.  Is this indeed the case for mixed entangled 
states, or is there some other, perhaps operational, characteristic 
that would better define the boundary between quantum and classical 
mechanics?  For example, we can make states that are mixed and 
non-separable and yet do not violate a Bell's inequality - are these 
``truly'' entangled?  Distilling these states makes states that are 
more mixed and more entangled and so that they now violate a Bell's 
inequality \cite{Distill} - was this entanglement really ``hidden''?  
Using the Tangle criteria, the answer is straightforward: the states 
were always entangled and the distillation simply moved their position 
on the Tangle-Entropy plane across the Bell boundary \cite{MunroBell}.  
Yet, states that do not violate Bell's inequality can be described by 
a hidden local variable model - suggesting that some entangled states 
are classical, in violation of Schr\"{o}dinger's precept.  The 
question of what significant physical differences, if any, exist 
between these various mixed entangled states remains open.

\begin{figure}[ht!]
\begin{center}
\epsfxsize=\columnwidth
\epsfbox{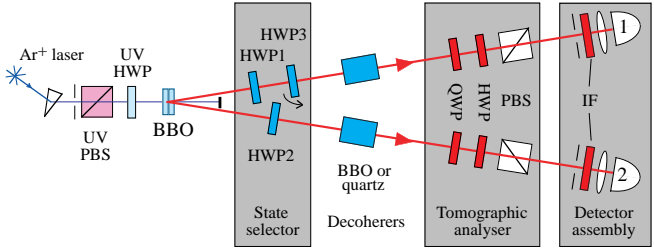}
\end{center}
\footnotesize \caption{Experimental setup for quantum state 
synthesis.}
\end{figure}
\begin{figure}[ht!]
\begin{center}
\epsfxsize=\columnwidth
\epsfbox{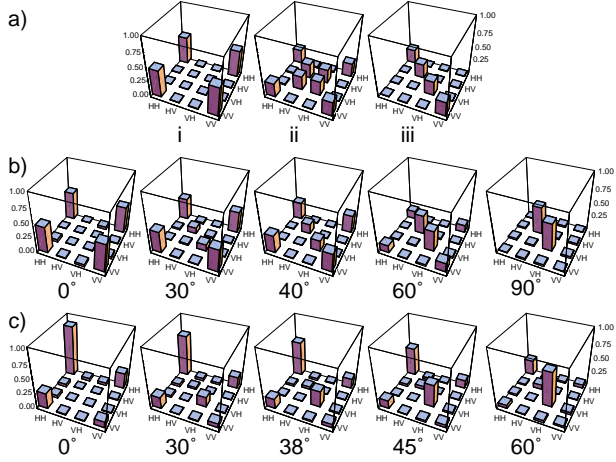}
\end{center}
\footnotesize \caption{{\it States obtained by spatially-based 
polarisation decoherence.\/} a) i) Input state 
$(|\rm{HH}\rangle+|\rm{VV}\rangle)/\sqrt{2}$; ii), iii) States after 
passing through BBO decoherers in one and both arms, respectively.  
{\it States obtained by temporally-based polarisation decoherence.\/} 
b) The state, $(|\rm{HH}\rangle+|\rm{VV}\rangle)/\sqrt{2}$, after 
passing through the state selector ($\theta_{1}=0$, $\theta_{2}$ set 
as shown) and the quartz decoherers set as described in the text.  c) 
As for b), but with the non-maximally entangled initial state 
$0.96|\rm{HH}\rangle+0.29|\rm{VV}\rangle$.  Only the real components 
of the density matrices are shown, the imaginary components being at 
the few percent level or less.}
\end{figure}
\begin{figure}[ht!]
\begin{center}
\epsfxsize=\columnwidth
\epsfbox{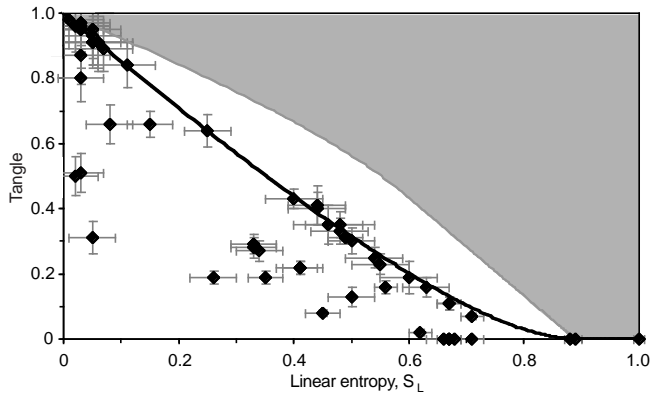}
\end{center}
\footnotesize \caption{Tangle vs linear entropy for 2 qubits.  {\it 
black curve\/}: Werner states.  Data points are calculated tangle and linear 
entropy from a range of measured density matrices.  The grey region 
indicates physically impossible combinations of T and $\rm{S_{L}}$; 
maximal states (Eq.  1) lie at the boundary of this region.}
\end{figure}
\newpage
\end{multicols}


\begin{references}
\bibitem{footnote} The tangle, T, is directly related to the 
Entanglement of Formation, E, which is regarded as the canonical 
measure of entanglement \cite{Tangle1,Tangle2}.  (Specifically, 
$\rm{E} = h\left(\frac{1+\sqrt{1-T}}{2}\right)$, where $h(x) =-x {\rm 
log}_{2}x-(1-x) {\rm log}_{2}(1-x)$.) For our 2-qubit system we use 
a normalised form of the linear entropy, $\rm{S_{L}}= 4(1-\rm{P})/3$, 
where the state purity is $\rm{P} = \rm{Tr}(\rho^{2})$ 
\cite{BoseVedral}. Note that $0 \leq \rm{S_{L}}, \rm{T} \leq 1$.
\bibitem{comment} By looking in coincidence, we restrict ourselves to 
the two-(and more)-photon section of the Hilbert space, i.e. we can 
disregard the vacuum component.  Two-photon states measured in this 
manner have been shown to near maximally violate Bell's inequality 
\cite{Weihs}, indicating that states are near-pure and 
near-maximally-entangled, i.e. strongly non-classical.
\bibitem{comment2} We stress that this principle - tracing over 
another degree of freedom in a tightly controlled way to induce 
decoherence - is applicable well beyond our specific experimental 
system.
\end{references}
\end{document}